\documentclass[fleqn,10pt]{wlscirep}
\usepackage[utf8]{inputenc}
\usepackage[T1]{fontenc}
\usepackage{graphicx,graphics}
\usepackage{amsmath}
\usepackage{amssymb}
\usepackage{epstopdf}
\usepackage{amstext}
\usepackage{amsthm}
\usepackage{amsfonts}
\usepackage{latexsym}
\usepackage{array}
\usepackage{xfrac}
\usepackage{color}
\usepackage[toc,page]{appendix}
\usepackage{subfigure}
\newcommand{\igbjd}[1]{}
\newcommand{\beqa}{\begin{eqnarray}}
\newcommand{\eeqa}{\end{eqnarray}}
\newcommand{\beq}{\begin{equation}}
\newcommand{\eeq}{\end{equation}}
\title{Entanglement structure of a quantum simulator: the two-component Bose-Hubbard model}

\author[1,2,*]{I. Morera}
\author[1,2]{Artur Polls}
\author[1,2,3]{Bruno Juli\'{a}-D\'{i}az}
\affil[1]{Departament de F\'isica Qu\`antica i Astrof\'isica, 
Facultat de F\'{\i}sica, Universitat de Barcelona, E--08028
Barcelona, Spain}
\affil[2]{Institut de Ci\`encies del Cosmos, Universitat de 
Barcelona, ICCUB, Mart\'i i Franqu\`es 1, Barcelona 08028, Spain}
\affil[3]{ICFO-Institut de Ciencies Fotoniques, The Barcelona Institute 
of Science and Technology, 08860 Castelldefels (Barcelona), Spain}

\affil[*]{imorera@icc.ub.edu}

\begin{abstract}
We consider a quantum simulator of the Heisenberg chain with ferromagnetic interactions 
based on the two-component 1D Bose-Hubbard model at filling equal to two in the 
strong coupling regime. The entanglement properties of the ground state 
are compared between the original spin model and the quantum simulator 
as the interspecies interaction approaches the intraspecies one. A numerical 
study of the entanglement properties of the quantum simulator state 
is supplemented with analytical expressions derived from the simulated 
Hamiltonian. At the isotropic point, the entanglement 
properties of the simulated system are not properly predicted by the 
quantum simulator.
\end{abstract}
\begin{document}

\flushbottom
\maketitle
%
%
\thispagestyle{empty}

\section*{Introduction}
The Bose-Hubbard model is 
nowadays almost ubiquitous in the interpretation of ultracold atomic gases  
experiments with optical lattices~\cite{lewenstein2012ultracold}. It 
provides the prime ingredient that 
allows ultracold atomic setups to mimic well-known many-body 
problems~\cite{RevModPhys.80.885,lewenstein2012ultracold}. In particular, it 
makes these systems extremely competitive for building quantum simulators of a 
wide range of notably difficult physical problems~\cite{Bloch2012,Gross995}. 
A particularly relevant example is the use of a two-component 
Bose-Hubbard (TCBH) model as a quantum simulator of spin 
models~\cite{PhysRevLett.90.100401,1367-2630-5-1-113,Gross995}. 
As pointed out in those papers different spins, e.g. $1/2$, 1, etc, 
can be simulated depending on the filling factor of the two species in 
the chain. In this article we concentrate on a specific configuration of 
filling one for both species, i.e. equal number of atoms of both species in the 
chain which maps into a spin$-1$ system. In this case, the TCBH maps, using 
perturbation theory, into a Heisenberg model with ferromagnetic 
interactions~\cite{PhysRevLett.90.100401}. 

In this article we consider the question: To what extent does 
the quantum simulator exhibit similar entanglement properties than the 
simulated Hamiltonian? In particular, we focus on critical regimes where 
specific entanglement properties universally characterize the phase of 
the system. The analysis will be performed in the strongly interacting regime, 
where the interaction strength of both species is equal and much larger 
than the tunneling rate. We will study the entanglement properties of the 
system as the interspecies interaction is increased towards the point where 
all interactions match. In this way, the simulated spin model goes from 
an anisotropic Heisenberg model into the Heisenberg isotropic one. Analytical 
results using perturbation theory will be complemented with numerical calculations 
using DMRG (density matrix renormalization group). In this way we can compare 
the entanglement present in the TCBH with that of the spin model, paying 
particular attention to the critical phases which appear in the latter.

\section*{Model}
We consider two bosonic species with contact-like 
interactions in a 1D optical lattice at zero temperature, described by the 
Bose-Hubbard Hamiltonian, 
\begin{equation}
H=-t \, \sum_i \sum_{\alpha=A,B} \left( \hat{b}_{i,\alpha}^{\dagger}\hat{b}_{i+1,\alpha}+\text{h.c.}\right) 
+\frac{U}{2}\sum_i \sum_{\alpha=A,B} \left(\hat{n}_{i,\alpha}\left(\hat{n}_{i,\alpha}-1\right)\right)
+U_{AB}\sum_{i}\hat{n}_{iA}\hat{n}_{iB} \,,
\label{eq:TwoBH}
\end{equation}
where $\hat{b}_{i\alpha}$ ($\hat{b}_{i\alpha}^{\dagger}$) are the annihilation (creation) 
bosonic operators at site $i=1,\dots,L$ for species $\alpha=A,B$, respectively, and 
$\hat{n}_{i\alpha}$ are their corresponding number operators. We have assumed equal 
tunneling strength, $t>0$, and repulsive intra-interaction strength, $U>0$, for both 
components. For the rest of the work we set the energy scale to $t=1$.
The ground state (GS) of Eq.~\eqref{eq:TwoBH} in the strong-coupling regime 
($U\gg t$) is a Mott insulator (MI) with a total filling 
$\nu=N_A/L+N_B/L\equiv\nu_A+\nu_B$. In this work we fix $\nu_A=\nu_B=1$. 

We define entanglement properties through the reduced density matrix obtained 
tracing out the right half of the system $\rho_{L/2}=\text{Tr}_R |\psi\rangle\langle \psi|$, 
where $|\psi\rangle$ is the ground state of the Hamiltonian~\eqref{eq:TwoBH}. The 
amount of entanglement is quantified with the von Neumann entropy 
$S_E=-\text{Tr}\rho_{L/2} \log \rho_{L/2}$. Finally, the entanglement spectrum 
(ES)~\cite{PhysRevLett.101.010504} is defined in terms of $\xi_i=-\log \lambda_i$, 
where $\lambda_i$ are the eigenvalues of the reduced density matrix. 

\section*{Results}
\begin{figure}[t]
\begin{subfigure}
\centering
\includegraphics[width=0.50 \columnwidth]{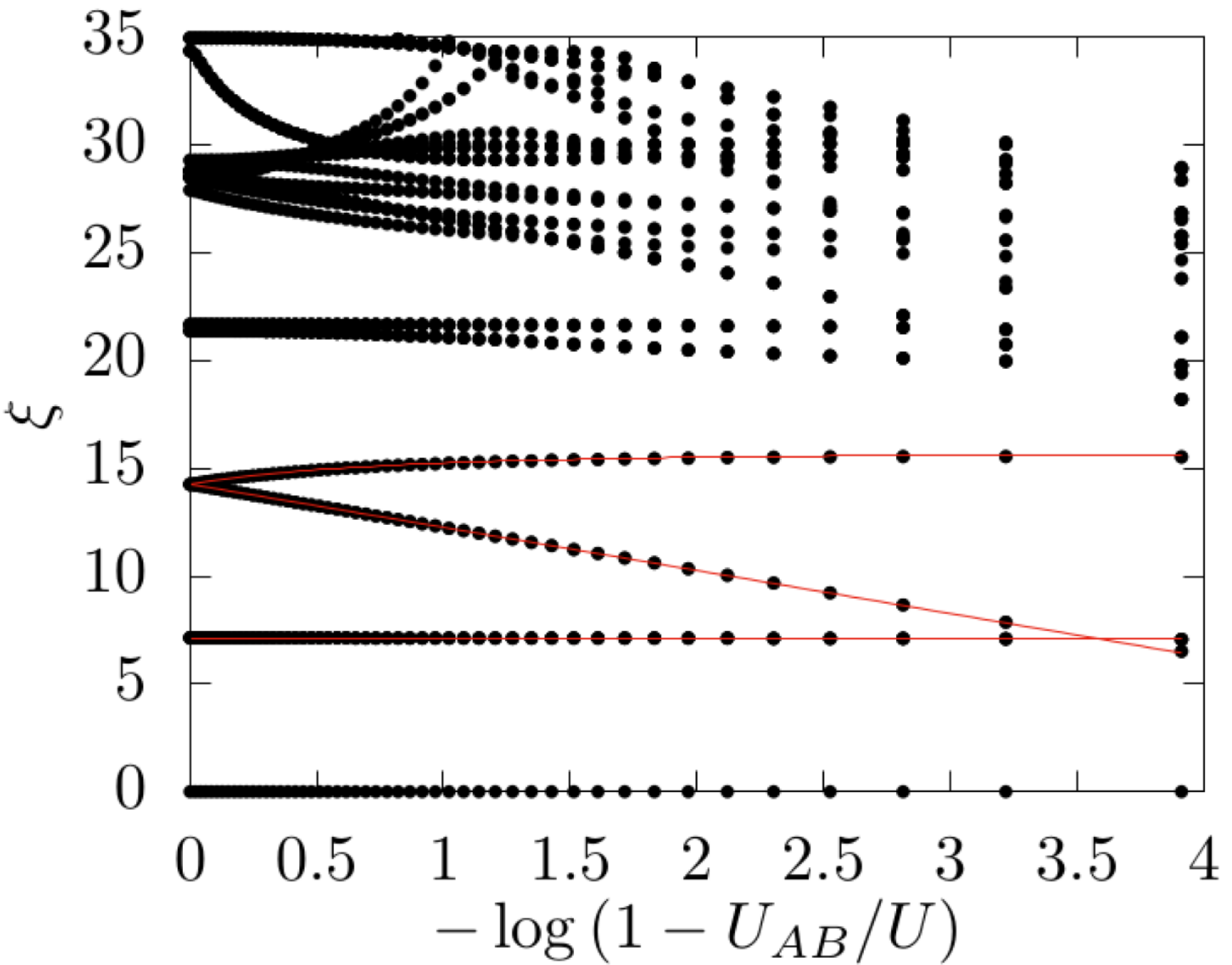}
\end{subfigure}
\begin{subfigure}
\centering
\includegraphics[width=0.50 \columnwidth]{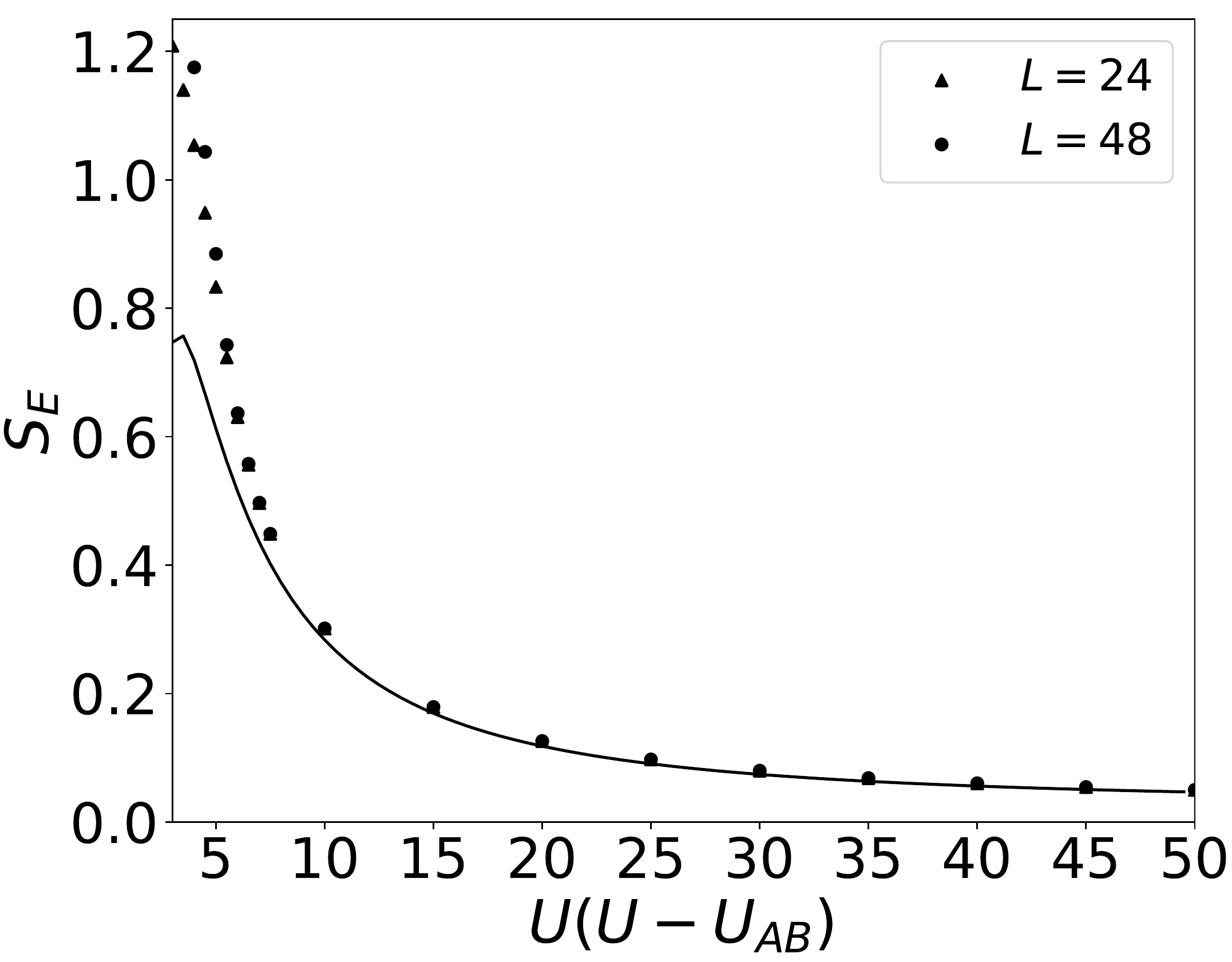}
\end{subfigure}
\caption{(Left panel) Entanglement spectrum of the TCBH (black dots) at fixed total size $L = 64$ 
as a function of $U_{AB}$ for fixed value of $U = 50$ obtained with DMRG. Continuous red 
lines represent analytical results Eqs.~\eqref{eq:ES_Analyt1} and~\eqref{eq:ES_Analyt2}, 
see text. (Right panel) von Neumann entropy $S_E$ as a function of $U_{AB}$ for fixed $U = 50$ 
and for different system sizes $L$. The Solid black line is the analytical result obtained 
through perturbation theory.}
\label{fig:spectrum}
\end{figure}

{\bf Perturbative regime.}
In the strong-coupling regime ($U\gg t$) the ES can be obtained perturbatively 
following ~\cite{PhysRevLett.108.227201}. In order to organize the ES we introduce 
the quantum numbers $\delta N_{\alpha} = N_{\alpha,L/2}-L/2$ which measure the excess 
($\delta N_{\alpha}>0$) or absence ($\delta N_{\alpha}<0$) of bosons ${\alpha} = A, B$ 
with respect to the MI with filling $\nu_A = \nu_B = 1$, on the left subsystem which 
is of size $L/2$. In Fig.~\ref{fig:spectrum} we report the obtained entanglement 
spectrum as a function of the interspecies interaction, $U_{AB}$, for a fixed, large, 
value of $U=50$. For $U_{AB}=0$ the structure of a single-component Bose-Hubbard (SCBH) 
model is recovered where different entanglement values are separated proportionally 
to the power $n$ of the perturbative parameter $1/U^n$, but for non-zero values 
$U_{AB}>0$ some entanglement values exhibit an explicit dependence on this 
interaction. The entanglement values associated with $\delta N_A=\pm 1$; 
$\delta N_B=0$ and $\delta N_A=0$; $\delta N_B=\pm 1$ are given by,
\begin{align}
&\xi^{(2)}_1= 2\log U - \log 2\,,
\label{eq:ES_Analyt1}
\end{align}
and do not show an explicit dependence on $U_{AB}$ at the order studied. Furthermore, 
these ones are completely analogous to the first ones obtained for the SCBH. The lowest 
entanglement value associated with $\delta N_A=\delta N_B=0$ gets a contribution 
$\xi^{(2)}_0=8/U^2$ due to the renormalization of the wavefunction.

Genuine second order contributions are of two different kind: $(+)$, the ones with 
$\delta N_A=\delta N_B$ which favor the movement of two different bosons through 
the boundary in the same direction and, $(-)$, the ones with $\delta N_A=-\delta N_B$ 
which favor the hopping of two different bosons through the boundary in opposite 
directions. Unlike $\xi_1^{(2)}$ these ones are totally absent in the SCBH as they 
are directly related to the presence of two different components. Specifically, 
configurations with $\delta N_A=-\delta N_B$ are associated to the phase separation 
of the two components through the boundary. An analytic formula can also be obtained, 
\begin{align}
&\xi^{(2)}_{\pm}= 4\log \left(U\right) +2\log\left(1\pm U_{AB}/U\right)-\log 4.
\label{eq:ES_Analyt2}
\end{align}

The two different branches, $(+)$ and $(-)$, have a very different behavior as $U_{AB}$ 
is increased. The $(-)$ one is seen to decrease as $U_{AB}/U\to 1$, predicting a closing 
of the entanglement gap at $U\left(U-U_{AB}\right)=2$. The analytical predictions are 
in very nice agreement with the DMRG calculations, which also show a closing of the 
Schmidt gap~\cite{PhysRevLett.109.237208} for $U-U_{AB}\simeq 1/U$. Note also, that the 
structure of the ES changes dramatically as we approach this point, with higher order 
processes becoming comparable to the lowest entanglement value. These higher order processes 
also show a logarithmic dependence as found for $\xi^{(2)}_{-}$ with a slope that indicates 
the order of the perturbation theory at which they are found. At this point we expect 
to enter in a critical regime. Notice that the von Neumann entropy $S_E$ starts to increase 
as these entanglement states decrease in the ES, see Fig.~\ref{fig:spectrum} (right panel) 
and a dependence on the system size starts to appear.

\begin{figure}[t]
\includegraphics[width=0.50 \columnwidth]{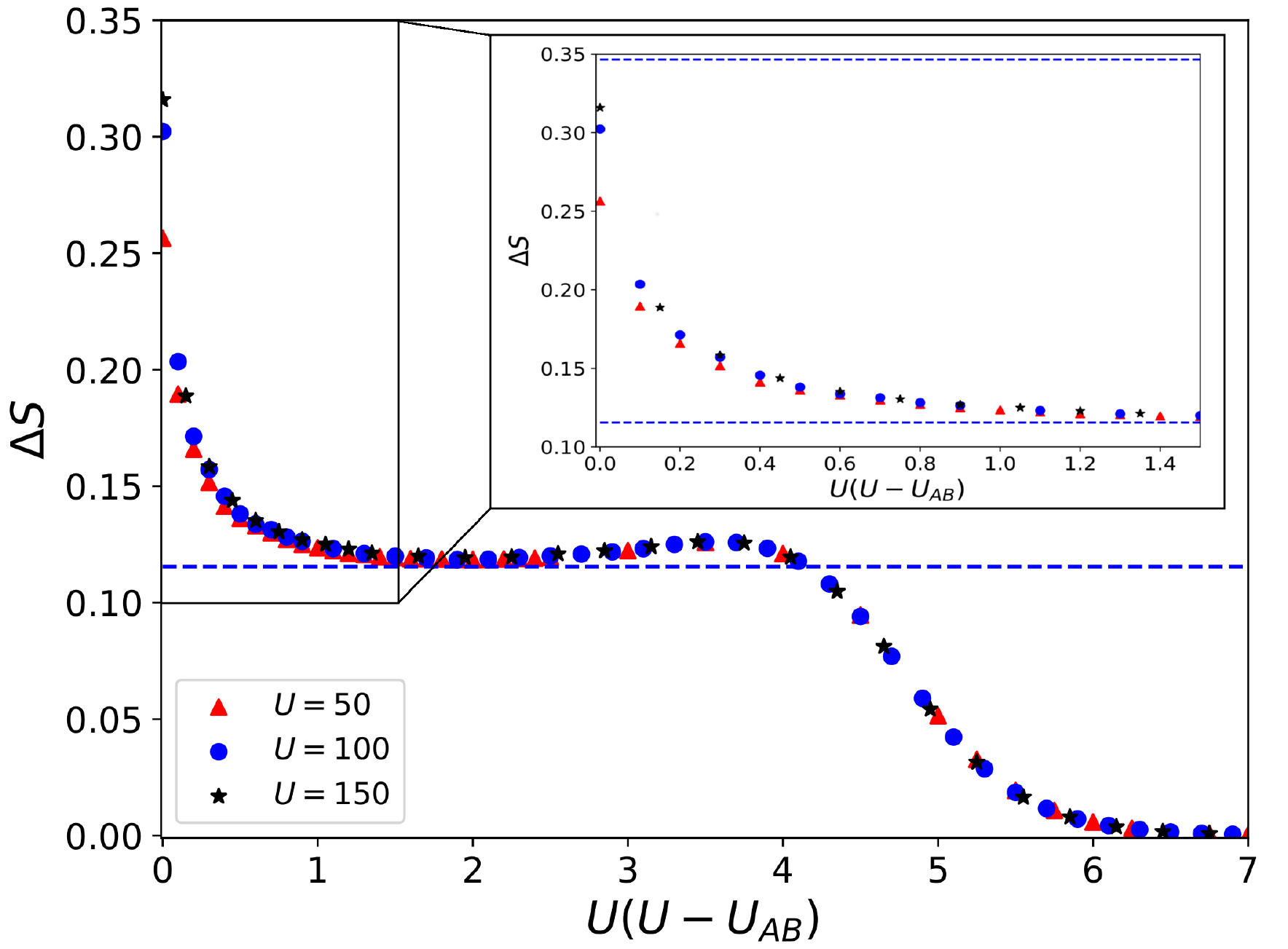} 
\caption{(Color online) Main panel: Entanglement scaling $\Delta S$ 
of the TCBH as a function 
of the universal coupling $U\left(U-U_{AB} \right)$ for different values of the interaction $U=50,100,150$ 
at fixed total system length $L=48$. The dashed line represents the CFT prediction for a theory with central 
charge $c=1$. Inset: Zoom in the region $U_{AB}\rightarrow U$. The upper continuous line represents the value 
predicted by the Heisenberg model and the lower one represents the CFT prediction for $c=1$. }
\label{Fig:Sscaling}
\end{figure}

{\bf The TCBH as a quantum simulator.} Interesting physics appears as we 
approach to $(U-U_{AB})\sim 1/U$. As discussed above, we enter in a critical regime 
which cannot be described by the simple perturbation theory. Instead one has to 
consider a degenerate perturbation theory. For the general case of $n$ 
atoms per site, the low-energy Hilbert space is described by an effective spin 
$S\equiv n/2$ where $A$ and $B$ are taken as a pseudo-spin $1/2$. The effective 
Hamiltonian describing this low-energy spin subspace is given by superexchange 
processes at second-order in the hopping~\cite{ANDERSON196399,PhysRevLett.90.100401,1367-2630-5-1-113}
\begin{equation}
H_{\rm eff}=-J\sum_i \mathbf{S}_i \mathbf{S}_{i+1} + D \sum_i \left( S_i^z \right)^2,
\label{eq:Heff}
\end{equation}
where $J=-4 t ^2/U$ and $D=U-U_{AB}$. Working with a fixed number of total bosons 
$\nu_A=\nu_B=1$ maps in the spin picture to the sector with null total magnetization 
in the z-axis $\sum_i S_i^z=0$ and an on-site total spin $S=1$. 

The model~\eqref{eq:Heff} has been extensively studied~\cite{PhysRevB.28.3914,PhysRevB.30.215,PhysRevB.34.6372,0953-8984-2-31-011,DegliEspostiBoschi2003,PhysRevB.67.104401} 
and presents different phases depending on the ratio $D/J$. Here, we consider $D \geq 0$. For 
$D/J \rightarrow \infty$ (large$-D$ phase) all spins tend to be in the zero $z-$projection and 
performing a perturbation calculation over this ground 
state at first order in $J$ leads to the 
same entanglement value $\xi_-^{(2)}$ previously found for the TCBH. At $D/J\ \sim 1$ the system  
enters in a critical XY ferromagnetic phase characterized by a conformal field theory (CFT) 
with central charge $c=1$~\cite{PhysRevB.34.6372,DegliEspostiBoschi2003}. Finally, 
for $D=0$ the system is in the isotropic point where its properties are governed by the $SU(2)$ 
symmetry of the Hamiltonian~\eqref{eq:Heff}. 

The equivalence between the Hamiltonians in Eq.~\eqref{eq:TwoBH} and Eq.~\eqref{eq:Heff} at a specific order in perturbation theory is what allows one to term the TCBH a quantum simulator of the Heisenberg model. 
But what happens with observables? We deal with this question by using Brillouin-Wigner perturbation 
theory introducing the wave operator $\Omega$. This $\Omega$ operator defines a mapping between 
the eigenfunctions in the subspace of the simulated Hamiltonian~\eqref{eq:Heff} 
$|\psi_0^{(n-1)}\rangle$ (obtained at order $n$ in perturbation theory) and the complete 
eigenfunctions of the quantum simulator Hamiltonian~\eqref{eq:TwoBH} at the same order 
$n$: $|\psi^{(n-1)}\rangle=\Omega \, |\psi_0^{(n-1)}\rangle$, the wave operator $\Omega$ also 
admits an expansion at $n-1$ order. Once the mapping between eigenfunctions is established, 
the requirement that any observable should also give equivalent results in the two models 
defines a mapping between observables $\hat{O}=\Omega \hat{O}_0\Omega^{\dagger}$. This also 
includes operators which only act on a subsystem, therefore the reduced density matrix 
$\hat{\rho}_{L/2}$ associated to this subsystem for an eigenstate $|\psi^{(n-1)}\rangle$ 
can be expressed in terms of the eigenstates $|\psi_0^{(n-1)}\rangle$
\begin{equation}
\hat{\rho}^{(n)}_{L/2}=\text{Tr}_R\{ \Omega |\psi_0^{(n-1)}\rangle \langle\psi_0^{(n-1)}| \Omega^{\dagger} \}.
\label{Eq:rhoRelat}
\end{equation}
With this mapping the question of how well the entanglement properties are 
reproduced in a quantum simulator is rewritten as: can $\Omega$ introduce 
some extra structure which affects the universal entanglement properties?

{\bf Entanglement in the critical regime.} The scaling of the von Neumann entropy 
can be used to characterize the different phases of the system. From a CFT description this 
is a well-known result~\cite{HOLZHEY1994443,1742-5468-2004-06-P06002} and the magnitude 
$\Delta S=S_E(L)-S_{E}(L/2)$ captures the scaling behavior properly~\cite{1742-5468-2008-05-P05018}. 
Following the known behavior of the simulated model, Eq.~(\ref{eq:Heff}), one expects to 
go from $\Delta S \rightarrow 0$ in the large$-D$ phase, to $\Delta S= \left( c/6 \right)\log 2$ 
in the critical XY phase with $c=1$. This is exactly what is seen in Fig.~\ref{Fig:Sscaling}, where 
we observe the crossover between the two regimes in the TCBH as we vary $U(U-U_{AB})\sim D/J$.  
Furthermore, these results are mostly independent of $U$, for sufficiently large $U$, and cross 
the CFT prediction at $U(U-U_{AB})=4.2\pm0.1$. Therefore, we can conclude that 
the transition in the spin picture from a large$-D$ to a critical XY ferromagnetic phase 
is captured by the transition observed in the TCBH. On the other hand, as $U-U_{AB}\rightarrow 0$ 
a dependence on $U$ starts to appear. 

\begin{figure}[t]
\begin{subfigure}
\centering
\includegraphics[width=0.35\columnwidth]{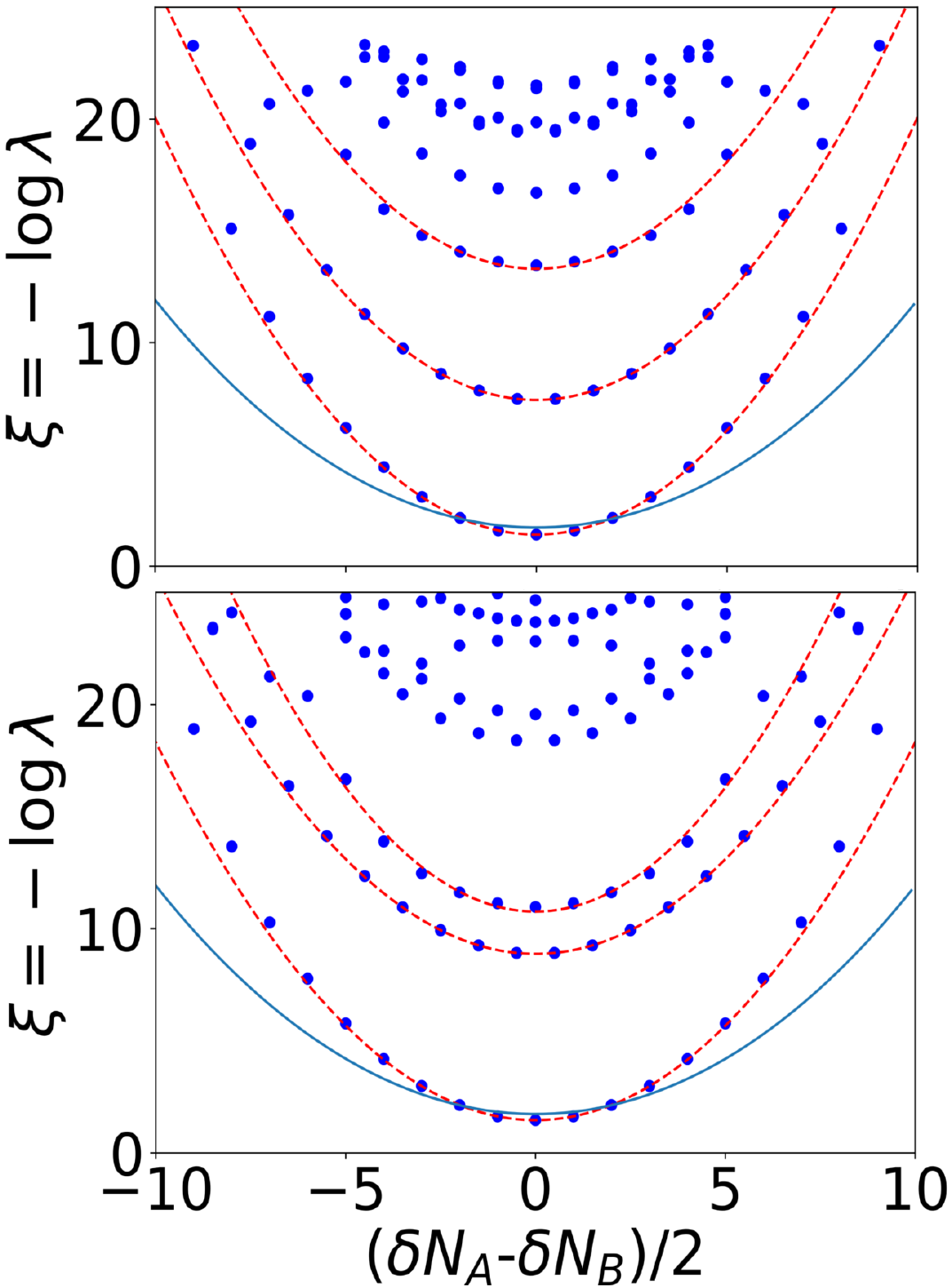}
\end{subfigure}
\begin{subfigure}
\centering
\includegraphics[width=0.60\columnwidth]{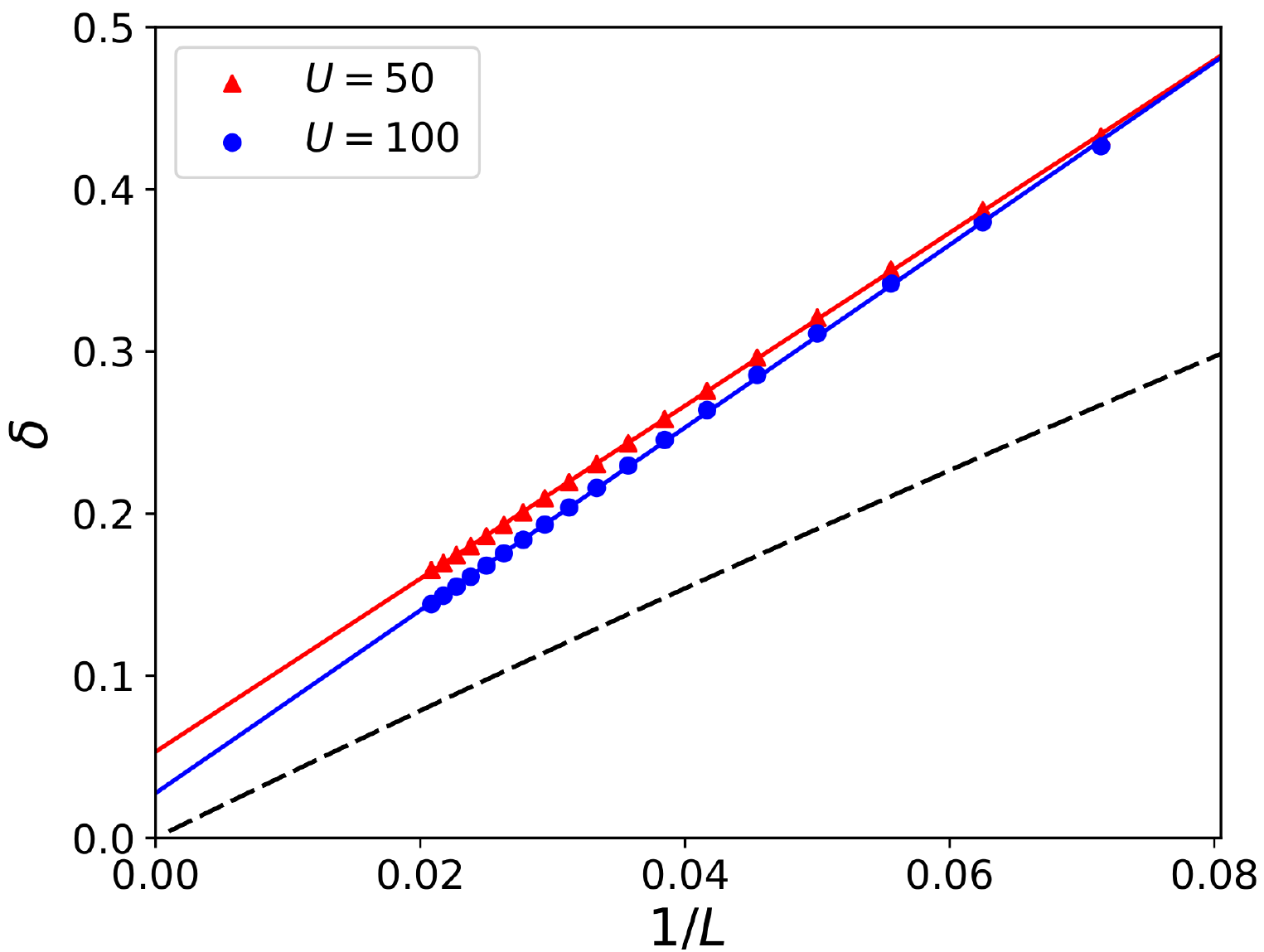}
\end{subfigure}
\caption{(Color online) Left panel: Entanglement spectrum of the TCBH model~\eqref{eq:TwoBH} 
at $U=50=U_{AB}$ (top) $U=100=U_{AB}$ (bottom) and $L=48$ as a function of the relative 
excess of bosons. Red dashed lines represent parabolic fittings and the continuous 
blue one is the analytical prediction given by the simulated spin model. Right 
panel: Entanglement gap (defined in the main text) as a function of the inverse of 
the system length $L$ for two different values of the interaction $U$ considering 
the critical point $U_{AB}=U$. Continuous lines represent linear fittings and the 
dashed one is the analytical value predicted by the simulated spin model.}
\label{Fig:Gap}
\end{figure}

{\bf Isotropic point.} In the spin model, the isotropic point $D=0$ is the 
end of the conformal line $c=1$ describing the critical XY phase, and the system 
only exhibits scale invariance~\cite{1742-5468-2011-02-P02001,PhysRevLett.108.120401}. 
The $SU(2)$ symmetry fully determines the ground state of the system, 
which is composed by a superposition of all the states belonging to the multiplet 
with maximum total spin. For a chain formed by $L$ spins $S$, this multiplet is obtained 
by applying the lowering operator $S^-=\sum_i S_i ^-$ to the fully polarized state 
$|S_T=SL,S_T^z=S L\rangle \equiv |F\rangle$. For specific sectors 
with fixed total magnetization $S^z _T\equiv SL-M$ the ground state of the system will 
be $|\psi_0\rangle = \left(S^-\right)^M|F\rangle$, which is a superposition of all spin 
configurations in the chain satisfying that the total magnetization is $S^z _T$. 
Therefore, considering a bipartition of the system $A$ of length $l$ the ES is 
organized by eigenstates with well defined magnetization $S^z_A=Sl-m$ in the subsystem 
$A$ with eigenvalues
\begin{equation}
\xi \left( m, M, S, L, l\right)=-\log \left( 
\frac{ \binom{2Sl}{m} \binom{2S \left( L-l\right)}{M-m}}{\binom{2SL}{M}} \right),
\label{Eq:ES}
\end{equation}
which is a natural extension of the results presented 
in~\cite{PhysRevA.71.012301,PhysRevE.82.011142,1742-5468-2011-02-P02001}.

From Eq.~\eqref{Eq:ES} an asymptotic expression for the von Neumann entropy 
$S_E$ can be obtained considering $l=L/2$ and $S^z_T=0$
\begin{equation}
S_E = \frac{1}{2}\log \left( \frac{SL \pi}{2} \right) + \frac{1 -  \log 2}{2}+\mathcal{O}(L^{-1}).
\label{Eq:Entro}
\end{equation} 
Notice that in the thermodynamic limit any small anisotropy $D>0$ will restore the 
conformal symmetry. For finite systems a smooth crossover between the CFT and the 
scale invariant prediction~\eqref{Eq:Entro} is expected~\cite{1742-5468-2013-10-P10007}, 
see inset Fig.~\ref{Fig:Sscaling}. In this region is where a non-universal behavior 
of the TCBH model is observed and we obtain different scalings of $S_E$ for different 
values of the interaction $U$. Furthermore, we observe that in the limit 
$\left( U-U_{AB} \right)\rightarrow 0$ the scaling mostly depends on the value 
of the interaction $U$ and does not coincide with the value predicted by the 
spin model, Eq.~(\ref{Eq:Entro}).

In order to understand the dependence of the scaling of the entanglement entropy 
on the interaction $U$ at $U=U_{AB}$ we examine the ES of the TCBH model~\eqref{eq:TwoBH} 
and compare it with the analytical prediction for the spin model~\eqref{Eq:ES}. The 
ES represented in Fig.~\ref{Fig:Gap} displays a parabolic dependence as a function 
of $\delta N_A-\delta N_B$ (which is analogous to $\delta S^z=S^z_T-S^z_A$ in the 
simulated spin model). This parabolic dependence is also expected from the spin 
picture Eq.~\eqref{Eq:ES} but the curvature is considerably different. Defining the 
entanglement gap $\delta = \xi^{(1)}-\xi^{(0)}$ as the gap between the two lowest 
entanglement values (which is directly related to the curvature of the parabola) 
one can observe that both depend linearly on the inverse of the system size 
$L$. But this linear dependence is different in the two models. Specifically, 
from the spin picture we obtain that $\delta \rightarrow 4/L$, so it closes in 
the thermodynamic limit $L\rightarrow \infty$. Conversely, in the TCBH model 
the gap does not close in the thermodynamic limit for finite values of the interaction. 
Furthermore, the ES predicted by the spin model~\eqref{Eq:ES} has a 
well defined magnetization $\delta S^z$, meaning that for each value of the magnetization 
there is a unique entanglement value $\xi_{\delta S_z}$. On the other hand, the ES of 
the TCBH model shows a richer structure with different parabolic envelopes for 
the same magnetization. Focusing on this extra structure we observe that the 
second parabolic envelope has associated a half-integer magnetization $\delta S_z$, 
unlike the first one which has integer magnetization. This can be understood 
expanding the wave operator at first order, $\Omega\simeq\left( 1-H_t/U\right)$, where $H_t$ is 
the hopping term of the Hamiltonian~\eqref{eq:TwoBH}. The second envelope is 
obtained by the application of $H_t|\psi_0^{(1)}\rangle$ over the frontier 
which defines the bipartition of the system for computing the ES. Therefore, 
these entanglement eigenstates correspond to having an extra particle or 
hole $\delta N=\pm1$ for any of the two species which explains the half-integer 
nature of $\delta S_z$. Notice that this component of the ground state 
wavefunction is a reminiscent of the first entanglement eigenstates with 
eigenvalue $\xi_1^{(2)}$, Eq.~(\ref{eq:ES_Analyt1}). But now because of the 
non-trivial entangled structure 
of the ground state $|\psi_0^{(1)}\rangle$ for each value of the subsystem 
magnetization we have this particle-hole excitation over the frontier which 
gives a large number of states, of order $L$. We have checked that the gap between 
the first two parabolic envelopes goes like $2\log U$ and does not show an explicit 
dependence on the system length $L$.

\begin{figure}[t!]
\includegraphics[width=0.50\columnwidth]{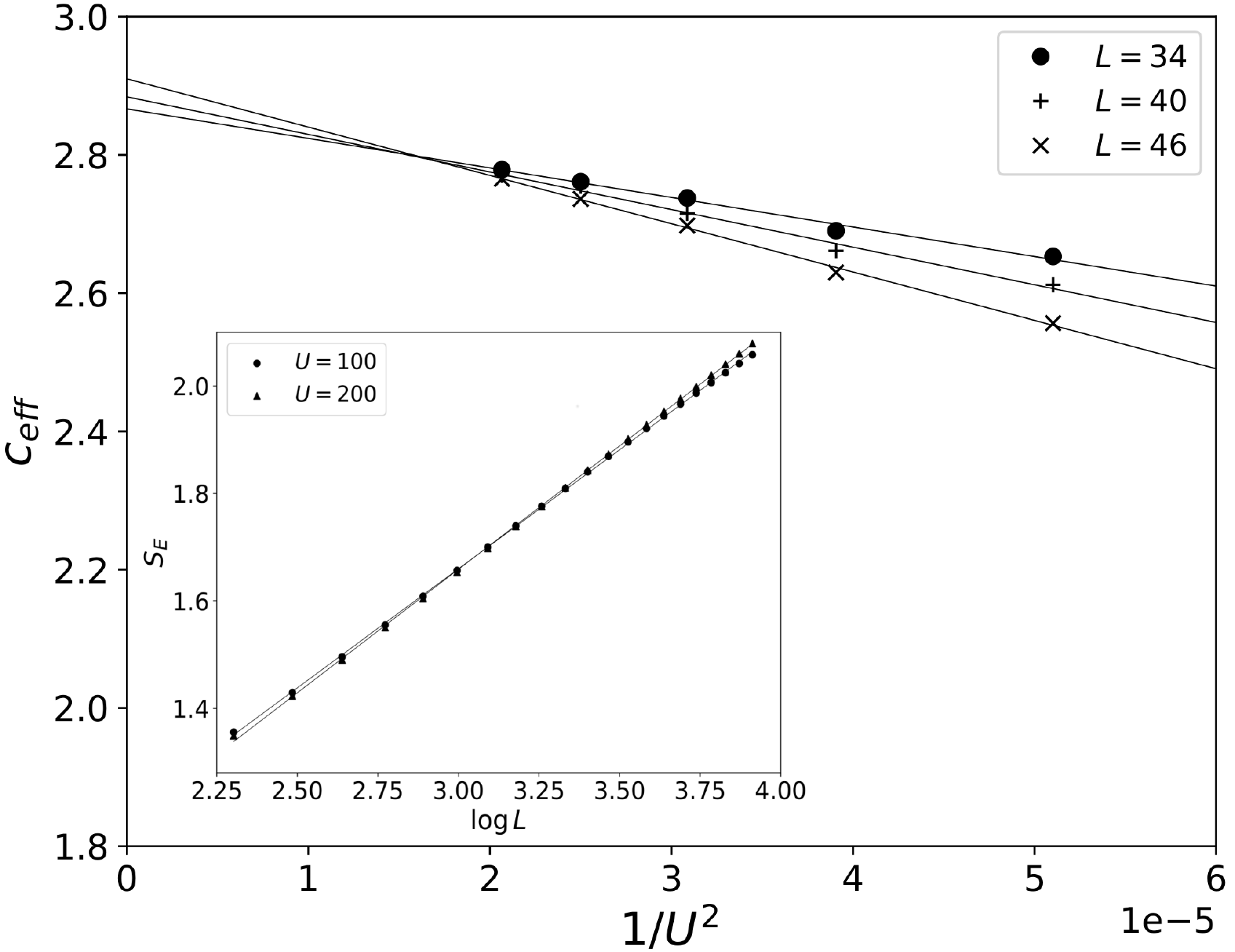}
\caption{(Color online) Main panel: Effective central charge (see main text) as a function of the 
inverse of the interaction $U$ considering $U_{AB}=U$ for different system sizes $L$. Inset: 
Entanglement entropy scaling for different values of the interaction. Continuous lines represent linear fittings.}
\label{Fig:C}
\end{figure} 

The effect of including $H_t|\psi_0^{(1)}\rangle$ in the wavefunction has 
large effects on the von Neumann entropy. The main reason is that the number 
of entanglement states given by $H_t|\psi_0^{(1)}\rangle$ is of order $L$ 
which is the same than the number of entanglement states in $|\psi_0^{(1)}\rangle$. 
Therefore, the contribution of both parts to the von Neumann entropy is 
$\log L$ and we can estimate the total contribution as 
$S_E\propto (1/2-A/U^2) \log L$, with $A$ some constant value. In order to 
verify that, we define the slope $c_{\rm eff}(L)=6\left(S_E(L)-S_E(L_0) \right)/(\log(L/L_0))$ 
with a reference size $L_0=50$, for which finite size effects will be 
reduced~\cite{PhysRevLett.109.267203}. In Fig.~\ref{Fig:C} we see that  
there is always a logarithmic behavior and the slope $c_{\rm eff}(L)$ shows 
a clear dependence on $1/U^2$ which confirms our predictions.

\section*{Conclusions}

The extent to which a quantum simulator of 
a well-known spin system captures the entanglement properties of the ground 
state of the simulated Hamiltonian has been scrutinized. We have considered 
the entanglement properties of the ground state of the two-component 1D 
Bose-Hubbard model in the strong-coupling regime for total filling 
$\nu_A=\nu_B=1$. This model acts as a quantum simulator of the spin 1 Heisenberg 
model with ferromagnetic interactions. In the regime in which the spin system 
is in a critical XY phase ($U-U_{AB}\sim t^2/U$) the two-component Bose-Hubbard 
model shows a universal (independent of the interaction $U$) scaling of the 
von Neumann entropy, which matches the CFT prediction expected for the 
simulated spin system. On the other hand, we observe that this universality 
is lost as we approach the isotropic point $U=U_{AB}$ where the simulated  
spin model loses the conformal invariance. By comparing the ES of the quantum simulator 
with the simulated spin model, which has been analytically obtained, we 
observe large discrepancies between the two of them for large values of 
the interaction $U$. In particular, magnitudes which should display a 
universal behavior (like the slope of the scaling in the entanglement 
entropy) strongly depend on the interaction $U$. This dependence has 
been analytically predicted constructing the wavefunction of the 
two-component Bose-Hubbard model using the wave operator.

\section*{Acknowledgements}
The authors thank V. Ahufinger, G. De Chiara, J.I. Latorre, and 
M. Lewenstein, J. Martorell and L. Tagliacozzo for useful comments 
and discussions. This work is partially funded by MINECO (Spain) 
Grant No. FIS2017-87534-P.

\section*{Author contributions statement}
I. M. and B. J.-D. conceived the idea, I. M. performed the calculations and all authors analyzed the results and wrote the manuscript.

\section*{Additional information}

\textbf{Competing interests:} The authors declare no competing interests.

\end{document}